\documentclass[11pt]{JHEP3}
\usepackage{epsfig,multicol,bbm}
\usepackage{cite}
\usepackage{amsmath,amsfonts,amssymb,longtable}
\usepackage{array} 
\usepackage{graphicx}
\usepackage{subfig}
\input epsf.sty

\def\be{\begin{equation}}
\def\ee{\end{equation}}
\def\ba{\begin{array}}
\def\bacc{\begin{array} {cc}}
\def\ea{\end{array}}
\def\bea{\begin{eqnarray}}
\def\eea{\end{eqnarray}}
\def\bd{\begin{displaymath}}
\def\ed{\end{displaymath}}
\def\calN{{\mathcal N}}

\def\Z{\mathbb Z}
\def\N{{\rm N}}
\def\L{{\rm L}}
\def\M{{\rm M}}
\def\R{{\rm R}}
\def\K{{\rm K}}
\def\k{{\bf k}}
\def\mod{{\,\,\rm mod}\,\,}
\def\nn{\nonumber}

\usepackage{cancel, color}

\title{\huge  Revisiting Coupling Selection Rules in Heterotic 
Orbifold Models}

\author{Tatsuo Kobayashi \!$^a$, Susha L.~Parameswaran$^b$, Sa\'ul Ramos-S\'anchez$^{c}$, Ivonne Zavala$^d$ \\ \\
${}^a$Department of Physics, Kyoto University, Kyoto 606-8502, Japan \\
${}^b$Department of Physics and Astronomy, Uppsala University, 
  P.O. Box 803, S-75108, Uppsala, Sweden\\ 
${}^c$Departamento de F{\'\i}sica Te\'orica, Instituto de F{\'\i}sica, UNAM, M\'exico D.F. 04510, M\'exico \\
${}^d$Bethe Center for Theoretical Physics and
  Physikalisches Institut der Universit\"at Bonn,
  Nussallee 12, 53115 Bonn, Germany\\ \\
E-mail: {\email{kobayash@gauge.scphys.kyoto-u.ac.jp}\,,
\email{susha.parameswaran@physics.uu.se}\,, 
\email{ramos@fisica.unam.mx}\,, 
\email{zavala@th.physik.uni-bonn.de}}
}

\preprint{KUNS-2341 \\ UUITP-18/11}

\abstract{
We study L-point couplings between twisted sector fields in heterotic
orbifold compactifications, using conformal field theory.  Selection
rules provide an easy way to identify which couplings are
non-vanishing.  Those used in the current literature are gauge
invariance, R-charge conservation and the space group selection rule,
but they are not the whole story.  We revive and refine a fourth
selection rule, due to symmetries in the underlying torus lattice, and
introduce a fifth one, due to the existence or not of classical
worldsheet instanton solutions to mediate the couplings.  We consider 
briefly the phenomenological consequences of the additional rules, in particular
for recent orbifold constructions whose field
content correspond to that of the MSSM.  
The structure of the exotic mass matrices is unaltered and many dimension-5 proton-decay
operators vanish.}


\begin{document}

\section{Introduction}
After around a quarter of a century of work, heterotic orbifold
compactifications continue to provide a promising framework for
building realistic models of Nature from string theory.  One of the
huge advantages of these scenarios is their simplicity.  Orbifold
compactifications are globally consistent constructions with a clear
geometrical interpretation, and thus several phenomena can be
understood in an intuitive geometrical way.  Moreover, 
they correspond to free CFTs, and in particular 
one can solve strings on the orbifold.
Hence, many quantities, such as the string couplings, can be computed exactly.  

Recently, it has become possible to build orbifold models whose
spectrum corresponds to the MSSM with no chiral exotics
\cite{buchmuller, buchmullerII, Saul}.
Also other interesting models have been constructed 
\cite{Kobayashi:2004ud,tatsuo-stuart,Kim:2006hw}.
  It is now important to study more closely the phenomenology of these
  models.  Among the issues that must be addressed are the decoupling
  of vector-like exotics, the hierarchy of quark and lepton masses and
  the suppression of proton decay operators.  To tackle these
  problems, the couplings in the low energy effective field theory are
  required.  The trilinear couplings including ground state twisted
  fields are well understood
  \cite{Hamidi:1986vh,Dixon:1986qv,Burwick:1990tu,Erler:1992gt,
    KobayashiOhtsubo,Kobayashi:1991rp}, and some studies including
  excited twisted states have also been made \cite{Hamidi:1986vh,BLS}.
  Analytic computations of higher order coupling strengths were
  considered recently in \cite{Choi:2007nb}.  In general, couplings between
  twisted strings are mediated by worldsheet instantons, 
and are exponentially suppressed in the area separating the participating states and wrapped by the instantons \cite{Hamidi:1986vh,Dixon:1986qv}.

Actually, the first question asked when considering couplings in a
given model is simply which ones are non-vanishing.  The non-vanishing
couplings are those allowed by the string theory selection rules.
These were discussed from the very beginning in the orbifold
literature \cite{Hamidi:1986vh,Dixon:1986qv}, and a comprehensive
account of their status in the eighties can be found in
\cite{Font:1988tp,Font:1988nc}.  Aside from the usual constraint of
gauge invariance, they may be  briefly summarized as follows.  (i)
(Point and) space group invariance.  This comes from the requirement
that the boundary conditions of participating strings are such that
they can interact. The various twisted sectors in a coupling
$\theta^{k_1} \theta^{k_2} \dots \theta^{k_\L}$ must satisfy $k_1 +
k_2 + \dots + k_\L = 0 \mod \N$ for a $\Z_\N$ orbifold with the twist
$\theta$, and the fixed points on which the twisted fields are localized are restricted.  (ii) H-momentum conservation.
The SO(10) lattice momentum associated with the bosonized right-moving fermions must be conserved.  (iii)  Twist invariance.  For factorizable orbifolds this means that the couplings are invariant under independent orbifold twists in each of the planes.  Thus, the numbers of bosonic oscillators in the corresponding correlation function are restricted.  (iv) Rule 4. When all twisted fields are at the same fixed point, the correlation function enjoys not only the twist symmetry but the full torus lattice symmetry, which can include an additional $\Z_2$ or $\Z_3$ symmetry.  In that case, the numbers of bosonic oscillators are restricted further.

These selection rules have evolved since their original formulation.
The H-momentum of a given state depends on the ghost-picture with
which we choose to write the corresponding vertex operator.  The same
holds for the number of right-moving bosonic oscillators.  Therefore,
H-momentum conservation and twist invariance have been incorporated
into a picture-independent R-charge conservation, corresponding to a
discrete R-symmetry in the low energy effective field theory
\cite{tatsuo-stuart}.   Attempts have also been made to understand the
stringy space group selection rule in terms of conventional global
symmetries in the field theory limit, with  partial success
\cite{tatsuo-stuart, buchmullerII, jonas}.
Also, these stringy selection rules can lead to non-Abelian discrete 
flavor symmetries \cite{tatsuo-stuart, Kobayashi:2006wq, Ko:2007dz}.
Meanwhile, Rule 4, which has also evaded a field theoretic interpretation \cite{Font:1988nc}, has been lost in the recent literature.  Thus, the current selection rules that have been applied for example in \cite{Saul} are gauge invariance, the space group selection rule and  R-charge conservation.  

The purpose of the present paper is to readdress the coupling selection rules in heterotic orbifold models in the light of the previous discussion.   In principle, the single condition of R-charge conservation is not sufficient to ensure both H-momentum 
and twist invariance.  We explain how R-charge conservation put together with point group invariance does however turn out to be sufficient.   We reinstate Rule 4, and study it in some detail.  We also identify a new selection rule, Rule 5, by which couplings may be forced to be vanishing if the classical instanton solutions that would mediate them are zero.
Our focus is on the derivation of the rules by considering the 
string theory CFT correlation functions,
and we leave their interpretation within the low energy effective field theory for future work.  By implementing Rules 4 and 5 in concrete models we confirm that they are non-trivial, and begin to explore their phenomenological implications.

The paper is organized as follows.  In the next section we briefly review heterotic orbifold constructions and in particular  the CFT ingredients necessary to compute couplings.  We also outline the standard selection rules.  Then we present the general form of the correlation function corresponding to L-point couplings in Section \ref{S:Lpoint}, and use this to derive 
H-momentum conservation and twist invariance, 
as well as Rules 4 and 5.  In Section \ref{S:Z6II} we illustrate how the new rules are implemented in the example of a $\Z_{6-II}$ orbifold, and begin to investigate their impact in explicit MSSM models.  Lastly, we close with some further discussion about the new rules.  

Comprehensive reviews on orbifold compactifications can be found in \cite{BLreview, choikim}, and accounts from a more recent perspective are given in \cite{patrick,saulsthesis,jonas}.

\section{Twist fields, vertex operators and correlation functions}

In this section we introduce the setup of heterotic strings on
orbifolds, and the basic ingredients necessary to compute couplings at
any order.  Having described the untwisted and twisted massless closed
string states on the orbifold, we construct the corresponding 
vertex operators in the conformal field theory.  The prescription for
computing the couplings is then to calculate the correlation functions
of the vertex operators, and integrate over the location of the
operators on the world-sheet.

\subsection{The building blocks}

We begin by taking a 6D torus, which is obtained as $\mathbb{R}^6/\Lambda$,
where $\Lambda$ is some 6D lattice that we usually classify using the
root lattices of the semi-simple Lie algebras of rank six.
To
construct the orbifold, we quotient the torus by a finite-order
automorphism of $\Lambda$, called the point group.  
Possible lattices $\Lambda$ to realize $\Z_\N$ orbifolds can be found in 
\cite{Katsuki:1989bf,Kobayashi:1991rp}.
For simplicity we
restrict ourselves to the $\Z_\N$ orbifolds whose underlying torus lattice can be factorized into three 2D torii.  
These so-called factorizable orbifolds are $T^6/\Z_3$, $T^6/\Z_4$, $T^6/\Z_{6-I}$ and
$T^6/\Z_{6-II}$.
The orbifold twist $\theta$ can be represented by the twist vector
\be
v = \frac{1}{\N} \left(0,0,a,b,c\right)\,,
\ee
 with the conditions $a+b+c=0 \mod \N$ 
and $a,b,c\neq 0 \mod \N$
required by $\calN=1$
supersymmetry.  We denote the components of the twist vector by $v^m$, with $m=1, \dots 5$.  The last three components describe the non-trivial
action of the twist on the three internal planes, and we call them $v^i$  
with $i=1,2,3$.

Heterotic string states on the orbifold are composed of the following
elements.  There are the bosonic strings on the 4D spacetime and the
6D orbifold, their right-moving superpartners and the left-moving
gauge parts. 
There are two kinds of massless closed string states
on the orbifold: untwisted strings and twisted ones.  For the
$\theta^k$-twisted sector ($0 \leq k \leq \N-1$ and $k=0$ is the untwisted sector), the
boundary conditions for the complexified 6D string 
coordinates, $X^i = X^{M= 2i-1}+iX^{M = 2i}$, are  
\be
X^i(\sigma+\pi) = (\theta^k X)^i(\sigma) + \lambda^i, \qquad \lambda^i \in
\Lambda^i \,, \label{twistbc} 
\ee
and we have corresponding boundary conditions on the remaining fields.
The center of mass of the twisted states are localized at the fixed
points or planes $f$ of the orbifold geometry, and we denote the latter
with their corresponding space group elements $(\theta^k, \lambda)$.
Thanks to the orbifold identifications, the fixed points $(\theta^k,
\lambda)$  and $(\theta^k, \theta^l\lambda +(1-\theta^k)\Lambda)$ for any integer $l$
belong to the same conjugacy class, and they are physically equivalent.   However,  the fixed points of $\theta^k$ (when $2 \leq k \leq N-2$) are not always fixed points of
$\theta$, so the conjugacy classes of a higher $\theta^k$-twisted
sector are not typically in one-to-one correspondence with the fixed
points of $\theta^k$.   In that case, physical states are
$\theta$-invariant linear combinations of states located at different
fixed points of $\theta^k$, 
which otherwise carry exactly the same quantum numbers 
\cite{KobayashiOhtsubo,patrick,saulsthesis}. 

The physical states in the string Hilbert space we just introduced are
equivalent to fields in the orbifold conformal field theory.  We
define a twist field $\sigma_{\left(k, f\right)}(z, \bar
z)$ as the field that creates a ground state in the sector twisted by
$\theta^k$ (for $1 \leq k \leq \N-1$) so: $|\sigma_{\left(k, f\right)} \rangle=\sigma_{\left(k,
  f\right)}(0,0) |0 \rangle$.  The twist field
incorporates the non-trivial boundary condition (\ref{twistbc}) by
inducing a non-trivial monodromy; near a twist field located at the
origin the field $X^i$ undergoes a phase rotation 
\be
X^i(e^{2\pi i} z, e^{-2\pi i }\bar z) = e^{2 \pi i {\k^i}} X^i(z, \bar z) ,
\ee 
where $z=e^{2(\tau + i \sigma)}$.  
This corresponds to a branch point with order $\k^i$ in each
plane, where $\k^i = k \, v^i \mod 1$, such that 
$0 < \k^i < 1$.
The operator product expansions (OPEs) for the free fields $\partial
X$, $\bar \partial X$, $\partial \bar X$ and $\bar \partial \bar X$ 
(where $\partial = \partial/\partial z$, $\bar \partial = \partial/\partial \bar z$)
with the twist fields in each plane are given by 
\cite{Dixon:1986qv}
\bea
&&\partial X(z,\bar z) \sigma(w,\bar w)
 \sim (z-w)^{-(1-\frac{k}{N})} \tau(w, \bar w) + \dots ,\cr 
&&\partial \bar X(z,\bar z)  \sigma(w,\bar w) \sim (z-w)^{-\frac{k}{N}}
\tau'(w, \bar w) + \dots ,\cr 
&&\bar \partial X(z,\bar z)  \sigma(w,\bar w) \sim (\bar z-\bar
w)^{-\frac{k}{N}} \tilde\tau'(w, \bar w) + \dots ,\cr 
&&\bar \partial\bar X(z,\bar z)  \sigma(w,\bar w) \sim (\bar z-\bar
w)^{-(1-\frac{k}{N})} \tilde\tau(w, \bar w) + \dots  \, , \label{tau}
\eea
where we have written down the most singular parts in the expansion.  Also,
we have defined four excited twist fields $\tau$, and the tildes denote
fields related by complex conjugation on the world-sheet. 

Having introduced the twist fields, we can write down the vertex
operators for twisted states.  For our purposes it is enough to consider
the zero 4D momentum limit.  The emission vertex for a twisted bosonic
field is then given by\footnote{We omit the cocycle factors needed to satisfy fermionic anti-commutation properties \cite{goddard}, which would determine the overall sign of the correlation functions.  
We also omit twist-dependent normalization factors and factors coming from the definition of the excited twist fields, Eq.~(\ref{tau}).}
\be
V_{-1} = e^{-\phi} \prod_{i=1}^3 (\partial X^i)^{{\cal N}_L^i} \,\,
(\partial \bar X^i)^{\bar{\cal N}_L^i} \,\, e^{i q_{sh}^{m} H^{m}}
\, e^{i p_{sh}^I X^I} \,
\sigma^{i}_{\left(k,f\right)} \,,
\ee
whilst for a twisted fermionic field it is
\be
V_{-1/2} = e^{-\phi/2} \prod_{i=1}^3 (\partial X^i)^{{\cal N}_L^i}
\,\, (\partial \bar X^i)^{\bar{\cal N}_L^i} \,\, e^{i q_{sh}^{(f)\,m}
  H^m} \, e^{i p_{sh}^I X^I} \,
\sigma^{i}_{\left(k,f\right)}  \,.
\ee
In these expressions, $H^{m}$ are the five, free bosonic fields
representing the right-moving fermions through bosonization, $X^I$ ($I=1,\dots,16$)
correspond to the gauge fields, and $\partial X^i$ and $\partial \bar
X^i$ denote bosonic oscillators for the left-movers. The number of oscillators
creating any excited massless modes are counted by the oscillator
numbers, ${\cal N}_L^i$ and $\bar{\cal N}_L^i$.
The momentum in the gauge part can be written as 
$p_{sh} = p +k V
+ n_{a} W_{a}$, with $p$ a vector in the $E_8 \times
E_8$ lattice, $V$ the shift vector that describes the embedding of
the twist in the gauge degrees of freedom and $n_{a}
W_{a}$ any discrete Wilson lines present.  Similarly, we have
written the so-called H-momentum carried by the bosonized fermions as  
$q_{sh}= q + k \, v $, where $q$  lies on
the $SO(10)$ (vector) weight lattice and $v$ is the twist vector. 
The H-momentum in the spinor representation is 
written as $q_{sh}^{(f)}$, and  is related to that in the
vector representation by $q_{sh} =  q_{sh}^{(f)} + (1,1,1,1,1)/2$.
Below, we will use the fact that the summation of H-charge over the
three internal planes is fixed for each vertex operator \cite{cvetic}.  In
particular, counting positive chiral states\footnote{Throughout the paper, we refer to the
  positive chiral states in the spectrum.  For their negative chiral
  partners, $V_{-1}$ has summed H-charge -1, and $V_{-1/2}$ has summed
 H-charge +1/2.}, $V_{-1}$ has summed H-charge +1, whereas $V_{-1/2}$ has
summed H-charge -1/2.  We have factored the bosonic twist fields into
three 2D components.  The final component of the vertex operators is a free scalar, $\phi$, related to the
superconformal ghost system, and the subscript on the vertex operator
$V$ indicates the ghost-charge.  Lastly,  we note that throughout the paper, we suppress an overall normalization in the vertex operators whose dimension is given  in terms of the string scale, $\alpha' = l_s^2$,  by $\alpha'^{1/2} \prod_{i=1}^3 {\alpha'}^{-\left({\cal N}_L^i - \bar{\cal N}_L^i \right)/2}$ \cite{polchinski} .  So with each field in a coupling, we have a suppression by one inverse mass dimension as expected.  

Note that untwisted fields with $k=0$ are included in our discussion, with the corresponding vertex operators obtained by taking $\sigma^i_{(k,f)}$ to 1 in the expressions above.  Untwisted fields are each associated to one of the planes, and we call $U_1, U_2$ and $U_3$, respectively, untwisted sectors with H-momenta 
$q_{sh}=(0,0,1,0,0), (0,0,0,1,0)$ and $(0,0,0,0,1)$.

It will be convenient to use the picture-changing formalism,
whereby physically equivalent vertex operators carry superconformal
ghost-charge that differ by an integer.  In particular we will use
bosonic twisted vertex operators in the 0-picture, which are given by
\bea
V_{0} &=&\left(\alpha'/2\right)^{\frac12} \sum_{j=1}^3 \left(e^{i q_{0}^{j\,m} H^m} \,
\bar\partial X^j+ e^{-i q_0^{j \,m} H^m} \,\bar\partial \bar X^j \right) \cr
&& \qquad \qquad \times  \prod_{i=1}^3
(\partial X^i)^{{\cal N}_L^i} \,\,
(\partial \bar X^i)^{\bar{\cal N}_L^i } \,\, e^{i q_{sh}^m H^m} \, 
e^{i p_{sh}^I X^I}\, \sigma^{i}_{\left(k,f\right)}\,.
\label{V0} 
\eea
The picture-changing operation\footnote{Notice that in our conventions, the fields transform under the orbifold twist as: $X^I \rightarrow X^I + 2\pi V^I$, $H^m \rightarrow H^m - 2\pi v^m$, $\partial X^i \rightarrow e^{i 2\pi v^i}\partial X^i$, $\partial \bar X^i \rightarrow e^{-i 2\pi v^i}\partial \bar X^i$, $\bar\partial X^i \rightarrow e^{i 2\pi v^i}\bar\partial X^i$ and $\bar\partial \bar X^i \rightarrow e^{-i 2\pi v^i}\bar\partial \bar X^i$, so the picture changing operator is twist invariant, as it must be.} 
to the 0-picture has a
contribution to the H-momentum given by 
$q_0^{1} = (0,0,1,0,0)$, $q_0^{2}= (0,0,0,1,0)$ and $q_0^{3} = 
(0,0,0,0,1)$, and introduces right-moving oscillators that 
will be counted by $\calN^i_R, \,\, \bar \calN^i_R$.   The $V_0$
vertex operator is thus a sum of terms, each with a summed H-charge of 
2 or 0.

With the vertex operators in hand, we can compute the scattering
amplitudes for the massless states and deduce the coupling terms in the
4D effective field theory.   A term $\Phi^{n+3}$ in the superpotential,
with $\Phi$ a chiral superfield with components $\left(\phi,
\psi\right)$, can be inferred  
most straightforwardly
from an interaction of the form $\psi \,
\psi \, \phi^{n+1}$. Therefore, we want to investigate tree-level\footnote{Recall that, as a result of holomorphicity, the superpotential does not receive corrections in string perturbation theory \cite{DS}.}
correlation functions of the form  
\be
\langle V_{-1/2}(z_1,\bar z_1) V_{-1/2}(z_2,\bar z_2)  V_{-1}(z_3,\bar
z_3)  V_{0}(z_4,\bar z_4) \dots  V_{0}(z_{n+3},\bar z_{n+3}) \rangle\,,
\ee
where the ghost-charges have been chosen to cancel
the background value of 2 on the sphere.  
Of course, if such a coupling is vanishing, then so must be its supersymmetric relatives, 
such as the terms $\phi^{n+2}\bar\phi^{n+2}$ in the scalar potential 
that arise from the superpotential.

In the following, we use the index $\alpha=1, \dots,\L$ to label the states
participating in an L-point coupling.

\subsection{The standard selection rules}
\label{sec:standardrules}

We have seen above that vertex operators consist of several parts: the
4D part (whose momentum we have set to zero), the 6D twist field, the
bosonized fermions, the gauge part and the left and right-moving
oscillators.  
Since we have a free field theory, the correlation functions also factor accordingly, with the parts corresponding to $H^m, X^I$ and ghosts given by the well-known result similar to the Veneziano amplitude.  They can be found for example in\cite{polchinski, Choi:2007nb}. 

Each part of the correlation function has its own selection rule for allowed
couplings.  The most familiar of these are conservation of 4D momentum and conservation of the momentum of the gauge part,  
\be 
\sum_{\alpha=1}^{\L} p_{sh\,\alpha}^I = 0 \,, 
\ee
which is simply the requirement of gauge invariance.  These conditions can be derived via the textbook result \cite{polchinski}:
\be
\langle \prod_{\alpha=1}^\L e^{i p_\alpha. X(z_\alpha)} \rangle \sim \delta^{(16)}\left(\sum_{\alpha=1}^\L p_\alpha \right) \prod_{\alpha,\beta=1, \alpha <\beta}^\L \left(z_\alpha-z_\beta \right)^{\alpha' p_\alpha.p_\beta/2} \,.
\ee

An analogous computation for the bosonized fermions' part of the correlation function implies that the total H-momentum in each plane\footnote{Throughout the paper we use the term ``summed H-momentum'' to refer to the sum of H-momentum over the three complex planes for a given state, and ``total H-momentum'' to refer to the total H-momentum in a given plane from all states participating in an interaction.}  must
also be conserved in an interaction. 
For example, for a 3-point coupling $\langle
V_{-1/2} V_{-1/2} V_{-1} \rangle$, we can express the rule in terms of
the H-momentum of the scalar components as  
\be
\sum_{\alpha=1}^{3} q_{sh \, \alpha}^i = 1\,,
\ee
where we have used the relation between the fermionic and bosonic representations $q_{sh}^i = q_{sh}^{(f)\, i}+ \frac12$.
 For the higher order couplings, we have to take care of the fact that
 the picture-changing operator gives non-trivial contributions to the
 H-momentum.  
We describe in the following section how H-momentum conservation then plays an important role in determining the structure of the corresponding correlation functions.  Meanwhile, a picture-independent conservation law that
turns out to
 incorporate H-momentum conservation can also be constructed, as we
 discuss below. 

Another invariance property of the correlation functions is twist invariance.  As we prove in the following section, for the factorizable orbifolds it happens that couplings are invariant under independent orbifold twists in each of the three 2D planes:
\bea
\partial X^i \rightarrow e^{i 2\pi v^i}\partial X^i, \qquad  
\partial \bar X^i \rightarrow e^{-i 2\pi v^i}\partial \bar X^i, \cr
\bar\partial X^i \rightarrow e^{i 2\pi v^i}\bar\partial X^i, \qquad  
\bar\partial \bar X^i \rightarrow e^{-i 2\pi v^i}\bar\partial \bar X^i \, .\label{twistdX}
\eea
This corresponds to a
constraint on the total oscillator numbers present in a coupling:
\be\label{twistinv}
{\cal N}_L^i + {\cal N}_R^i - \bar{\cal N}_L^i - \bar{\cal N}_R^i = 0 \mod \N^i\,,
\ee
with $\N^i$ the orders of the orbifold twist in the 2D plane, i.e. the
smallest possible integers such that $\N^i v^i  = 0 \mod 1$ (no
summation over $i$).   Also, here and below, we denote
$\calN_L^i = \sum_{\alpha=1}^\L \calN_{L \, \alpha}^i$ and so on. 

Picture-changing, as well as introducing new H-momentum, incorporates
new oscillators into the vertex operators and correlation functions.  However, for a given state we can
define the picture-invariant R-charge  \cite{tatsuo-stuart,Font:1988nc}
\be
\label{eq:Rcharges}
\R^i_\alpha = q_{sh \, \alpha}^i - {\cal N}^i_{L\,\alpha} + \bar{\cal N}^i_{L\,\alpha} \,,
\ee
for which the contributions from the additional H-momentum and right-moving oscillators in the picture-changing operator cancel against each other. 
Then we can define the R-charge
conservation law as  
\be
\label{eq:Rconservation}
\sum_{\alpha=1}^\L \R_{\alpha}^i = 1 \mod {\N}^i \,.
\ee
It can be seen as a consequence of combining H-momentum conservation and twist invariance.  

The final well-known selection rule arises from
the requirement that the boundary conditions of the twisted closed
strings are such that they can join together to form other closed
strings.  This is called the space group selection rule, and takes the
form 
\be
\prod_{\alpha=1}^\L \left[ g_{\alpha} \right] = (\mathbbm{1},0) \,,
\ee
where $ \left[ g_{\alpha} \right]$ represents some element of the conjugacy
class of the space group element $g_\alpha$.  The space group selection
rule includes the point group selection rule, which requires
$\prod_{\alpha=1}^\L \theta^{k_\alpha} = 1$, or $\sum_{\alpha=1}^\L
k_\alpha = 0 \mod \N$ for the $\Z_\N$ orbifold.   In terms of the explicit space group elements, 
it can be written as
\bea
\!\!\!\!\! &&\!\!\!\!\!(1-\theta^{k_\L})(\theta^{m_{\L}}f_\L + \tau_\L) + \theta^{k_\L}(1-\theta^{k_{\L-1}})({\theta^{m_{\L-1}}}f_{\L-1} + \tau_{\L-1}) 
  + \ldots \nn \\
 &&\hskip4cm \ldots + \theta^{k_{\L}+ k_{\L-1}+ \dots + k_2}(1-\theta^{k_1})({\theta^{m_{1}}}f_1 + \tau_1) = 0 \,, \label{SGSR}
\eea
for some $\tau_{\alpha} \in \Lambda$ and arbitrary $m_\alpha\in \mathbb{Z}$.  Thus we see 
that the space group selection rule restricts the combinations of fixed points 
that can enter a coupling.  

In summary, the selection rules that are applied in the current literature are gauge invariance, R-charge conservation and the space group selection rule.  We observe here that, in principle, the  R-charge conservation law is not a sufficient condition to ensure that the two constraints of twist invariance and H-momentum conservation are individually respected, but only a necessary one\footnote{We thank Nana Cabo-Bizet and Dami\'an Mayorga-Pe\~na for raising this point and for discussions on it.}.  However,  one can check in explicit models that all the couplings allowed by both point group invariance and R-charge conservation, automatically also satisfy H-momentum conservation.  Thus, imposing point group invariance and R-charge conservation is  in fact enough to ensure that both the H-momentum and twist invariance conditions are satisfied.   We shall discuss this in a little more detail in the following section.

\section{L-point correlation functions and more selection rules} \label{S:Lpoint}

In the previous section we outlined the basic building blocks
necessary to study correlation functions for twisted states in heterotic orbifolds.  We
now study in more detail the structure of the correlation
functions.  We assume that the standard selection rules of gauge
invariance, R-charge conservation and the space group selection rule
have been applied, and identify two further rules that force some
couplings to be vanishing.

Our starting point will be
the non-trivial part of the general correlation
function, which takes the form (we set $\alpha'=2$)
\be
{\mathcal F} = \prod_{i=1}^3 \langle (\partial X^i)^{{\cal N}_L^i}
\,\, (\partial \bar X^i)^{\bar{\cal N}_L^i}  \,\, (\bar\partial
\bar X^i)^{\bar{\cal N}_R^i}  
\sigma_{\left(k_1,f_1\right)}^i \cdots
\sigma_{\left(k_\L,f_\L\right)}^i 
\rangle \, , \label{beforesplit}
\ee
where we have factorized it into 2D components.

\subsection{H-momentum conservation}

In writing the above expression, we have applied the H-momentum conservation law, 
which has the following consequences.  
Firstly, since the H-momentum in each plane must be
conserved, so must be their sum.   But recall that the summed 
H-charge for each vertex operator is fixed, in particular, $V_{-1}$
has charge +1, whereas $V_{-1/2}$ has charge -1/2.  Subsequently,  in the correlation function $\langle V_{-1} V_{-1/2}
V_{-1/2} V_0...V_0 \rangle$ only the terms with zero summed H-momentum
in $V_0$ contribute.  
These are the $\bar\partial \bar X^j$ terms in (\ref{V0}).  
Therefore, we have ${\cal N}_R^i =0$ and $\sum_i
\bar{\cal N}_R^i = \L-3$ \cite{Font:1988tp,Font:1988nc,cvetic}.  

Moreover, the H-momentum conservation in each plane reduces the correlation 
function to a form that factorizes over the three 2D directions, and determines how the right-moving oscillators are distributed amongst the three planes.  
In detail, although the correlation function is a sum of several terms due to the sums in the picture-changing operators of (\ref{V0}), all these terms are vanishing unless they satisfy the H-momentum conservation due to the result:
\be
\langle \prod_{\alpha=1}^\L e^{i q_\alpha. H(\bar z_\alpha)} \rangle \sim \delta^{(5)}\left(\sum_{\alpha=1}^\L q_\alpha \right) \prod_{\alpha,\beta=1, \alpha <\beta}^\L \left(\bar z_\alpha-\bar z_\beta \right)^{\alpha' q_\alpha.q_\beta/2} \,. \label{H-momentum}
\ee
This implies the condition 
\be
\sum_{\alpha} q_{sh \,\alpha}^i - \bar{\cal N}_R^i = 1, \label{Hmomentum}
\ee
for $\bar{\cal N}_R^i$, which satisfy $\bar{\cal N}_R^i \geq 0$ and $\sum_i
\bar{\cal N}_R^i = \L-3$.
As we commented above, it turns out that in explicit models the above conditions can be satisfied for all couplings that satisfy both the point group selection rule and R-charge conservation.  Indeed, it is easy to check that the couplings allowed by the point group selection rule,  which violate H-momentum conservation (see e.g. \cite{KobayashiOhtsubo} for a list of H-momenta for the various twisted sectors in $\Z_\N$ orbifolds),  also violate R-charge conservation.  This is because the planes in which H-momentum cannot be conserved turn out to be ones in which all fields are untwisted, with $q_{sh \, \alpha}^i=0$.  Moreover, in invariant planes massless modes do not have oscillators, so that R-charge reduces to H-momentum.  Then it follows that $\sum_{\alpha = 1}^\L \R_\alpha^i = 0 \neq 1 \mod \N^i$.  We shall see this in an explicit model in Section \ref{generalLpt}.

\subsection{Decomposition into classical and quantum parts}

To make further progress on the correlation function (\ref{beforesplit}), we split the computation into classical and
quantum parts.  Indeed, the fields $X^i$ can be split into a
classical instanton solution, which solves the equation of motion
$\partial \bar\partial X_{cl}^i = 0$, and the quantum fluctuations
about it: 
\be
X^i(z,\bar z) = X_{cl}^i(z, \bar z) + X_{qu}^i(z, \bar z) \,.
\ee
For a symmetric orbifold, which we will always assume, we have the
relations:
\bea
\bar\partial \bar X^i_{cl} &=& \left( \partial X^i_{cl} \right)^* ,\cr
\bar\partial X^i_{cl} &=& \left( \partial \bar X^i_{cl} \right)^* \, .
\eea
Of particular importance in what follows will be the explicit form for
the classical solutions, which describe worldsheet instantons.  The functional dependence of the classical
solutions is determined by the local monodromy to be \cite{Dixon:1986qv,Bershadsky:1986fv,Atick:1987kd}
\bea
\partial X^i_{cl}(z) &=& \sum_{l=1}^{\L-\M^i-1} a_{l}^i \,  h_{l}^i(z),
\label{dXcl}\\ 
\bar\partial X^i_{cl}(\bar z) &=&  \sum_{l'=1}^{\M^i-1} b_{l'}^i  
\, 
\bar {h'}_{l'}^{i}(\bar z) \,, \label{bardXcl}
\eea
plus their complex conjugates (no summation over $i$).  
Here, the basis functions are
\begin{eqnarray}
h_l^i(z) &=&  z^{l-1} \prod_{\alpha=1}^{\L}(z-z_\alpha)^{\k_\alpha^i -
  1}, \qquad l = 1,
\dots, \L-\M^i-1 ,\label{h}\\ 
\bar {h'}_{l'}^i(\bar z) &=&  \bar z^{l'-1}\prod_{\alpha=1}^\L (\bar z - \bar
z_\alpha)^{-\k_\alpha^i},  \qquad 
l' = 1, \dots, \M^i-1 \,,\label{h'} 
\end{eqnarray}
and the coefficients $a_{l}^i,b_{l'}^i$ shall be computed below (cf. eq.~\eqref{a,b}).
Also,  the  integers $\M^i$ are given by $\M^i=\sum_{\alpha=1}^\L
\k_\alpha^i$, and we have defined $\k_\alpha^i = k_\alpha  \, v^i \mod
1$, such that $0 < \k_\alpha^i \leq 1$  in (\ref{dXcl}) and (\ref{h}),
and $0 \leq \k_\alpha^i < 1$ in (\ref{bardXcl}) and (\ref{h'}).  The
integers $\M^i$ give the range of $l,l'$, and are determined from the
requirement that the classical solutions correspond to a
convergent classical action\footnote{Recall that the integral $\int dz
  \, d\bar z \prod_{\beta}|z-z_{\beta}|^{n_{\beta}}$ converges if and
  only if $n_{\beta} > -2$ for all $\beta$ and $\sum_{\beta} n_{\beta}
  < -2$.}. 
Indeed, only the classical solutions with finite action contribute 
in the path integral, while the solutions leading to divergent action 
have no contribution.
Then, the set  of holomorphic basis functions (\ref{h}) is non-empty when 
\begin{equation}
1 + \sum_\alpha (-1 + \k_\alpha^i) <0\,,
\end{equation}
whereas the set of anti-holomorphic basis functions (\ref{h'}) is non-empty when 
\begin{equation}
1 + \sum_\alpha (-\k_\alpha^i) < 0\,.
\end{equation}  
Otherwise, holomorphic and/or anti-holomorphic worldsheet instantons are not relevant,
and we must take instead the trivial solutions, respectively $\partial X_{cl}^i = 0$ and/or $\bar\partial X_{cl}^i=0$.

For any coupling involving only twisted fields
 in the plane $i$ ($\k_\alpha^i$ non-integers), the value of $\M^i$ 
lies between 1 and $\L-1$, and the total number of holomorphic plus anti-holomorphic 
functions is $\L-2$.
  The $\L-2$ coefficients, $a_l^i, b_{l'}^i$,
which complete the description of the classical solutions are
determined by the global monodromy conditions
(the quantum part instead feels only the local monodromy):     
\begin{eqnarray}
\int_{\gamma_p} dz \partial X^i_{cl} + \int_{\gamma_p} d\bar z
\bar\partial X^i_{cl} = \nu_p^i , \label{globmon} \cr 
\int_{\gamma_p} dz \partial \bar X^i_{cl} + \int_{\gamma_p} d\bar z
\bar\partial \bar 
X^i_{cl} = \bar{\nu}_p^i \,,
\end{eqnarray}
where $\gamma_p$ represent all possible net zero-twist closed loops enclosing the twist fields, 
and $\nu_p^i$  are elements of the corresponding cosets of the torus lattice. 
The number of independent equations arising from (\ref{globmon}) is
the number of  independent net zero-twist closed loops, which was
proven to be $\L-2$ in \cite{Atick:1987kd}. 
We can choose as a basis e.g. the loops used in \cite{Burwick:1990tu,Erler:1992gt}, where we encircle the
fixed point $f_p$ clockwise $r_p$ times followed by the fixed point $f_{p+1}$ counterclockwise $s_p$ times.  
Here, $r_p \, k_p = s_p \, k_{p+1} \mod \N$, with $r_p, \,s_p$ the smallest integers that satisfy this property, and $p=1, \dots, \L-2$.
The corresponding coset vectors, $\nu_p$, can then be
written as 
\begin{eqnarray}
\nu_p =
(1-\theta^{r_p \,k_p})(f_{p+1}-f_p+\lambda)\,, \qquad \lambda \in \Lambda \,.
\label{cosets} 
\end{eqnarray}
The global monodromy conditions thus stated are then just the required
number to determine all the $\L-2$ coefficients, $a_l^i, b_{l'}^i$, which reduces to an exercise in linear algebra.
Indeed, the solutions to (\ref{globmon}) can be written in terms of
the so-called period matrices, which have dimension $(\L-2) \times (\L-2)$
and components \cite{Atick:1987kd}
\begin{eqnarray}
W_{\,\,\,p}^{i\,l} &=& \int_{\gamma_p} dz h_l^i(z), \qquad l = 1,
\dots, \L-\M^i-1, \cr 
W_{\,\,\,p}^{i\, (\L-\M^i-1+l')} &=& \int_{\gamma_p} d\bar z \bar
{h'}_{l'}^i(\bar z), \qquad 
l' = 1, \dots, \M^i-1  \,.
\end{eqnarray} 
In terms of these matrices, the coefficients are
\begin{eqnarray}
a_l^i &=& \nu_p^i (W^{-1})^{i\,p}_{\,\,\,l} \,, \cr
b_{l'}^i &=&{\nu}_p^i (W^{-1})^{i\,p}_{\,\,\,L-M^i-1+l'} \,.
\label{a,b} 
\end{eqnarray}
 Thus for a particular classical instanton solution, the coefficients
 $a_l^i, b_{l'}^i$ are particular linear combinations of the coset
 vectors $\left\{\nu^i_p \right\}$. 

One more comment on the global monodromy conditions is in order here.
 In general, there is also an additional 
 consistency condition  \cite{Erler:1992gt} arising from the space
 group selection rule and the $(\L-1)th$ $(r_p,s_p)-$loop.  Indeed, the  $(\L-1)th$ loop is not linearly independent because the sum of all the $(\L-1)$ loops can be pulled around the worldsheet sphere and shrunk to zero,  giving rise to the space group selection rule.  
However, this extra consistency condition can restrict further the
 coset vectors appearing in (\ref{cosets}), (\ref{a,b})\footnote{This condition has not been
 considered in previous works on higher order couplings
 \cite{Abel:2003yx,Choi:2007nb}.}.  For instance, for the 3-point couplings it turns out that the coset vectors are restricted to \cite{Erler:1992gt}:
\be
\nu = \left(1-\theta^{r_1 \,k_1}\right) \left(f_2 - f_1  - \tau_2 + \tau_1 + (1-\theta^{k_1+k_2}){(1-\theta^{{\rm gcd}(k_1,k_2)})^{-1}} \lambda \right)\,,  \quad \lambda \in \Lambda\,,
\label{3ptcoset}
\ee
where $\tau_{1,2}$ are the torus lattice vectors that appear in the space group selection rule (\ref{SGSR}) and {\rm gcd} stands for greatest common denominator.

\bigskip

Having split the fields into their classical and quantum parts, the
 correlation function ${\mathcal F} = \prod_{i=1}^3 {\mathcal F}^i$
 also splits as  
\bea
&&{\mathcal F}^i = \sum_{r=0}^{ {\cal N}_L^i} {\scriptsize \begin{pmatrix} {\cal N}_L^i \\ r \end{pmatrix} }  \sum_{s=0}^{{\bar{\cal N}}_L^i}  {\scriptsize\begin{pmatrix} {\bar{\cal N}}_L^i \\ s \end{pmatrix}}\sum_{t=0}^{\bar{\cal N}_R^i} {\scriptsize\begin{pmatrix} \bar{\cal N}_R^i \\ t \end{pmatrix}} 
\sum_{X^i_{cl}} e^{-S^i_{cl}}  \times \label{Fqu+Fcl} \\
&& \qquad \qquad   \langle (\partial X^i_{cl})^{{\cal 
 N}_L^i-r}  \,\, (\partial X^i_{qu})^r \,\, (\partial \bar X^i_{cl})^{\bar{\cal N}_L^i-s}  \,\,(\partial \bar
 X^i_{qu})^s\,\, (\bar\partial \bar X^i_{cl})^{\bar{\cal N}_R^i-t}  \,\, (\bar\partial \bar X^i_{qu})^t  
 \rangle_{\sigma^i_{\left(k_1,f_1\right)} \cdots
 \sigma^i_{\left(k_\L,f_\L\right)}}\,,  \nonumber
\eea
where $\scriptsize{\begin{pmatrix} {\cal N}_L^i \\ r \end{pmatrix}}$ and so on are the binomial coefficients, the classical action is given by 
\be
S^i_{cl}= \frac{1}{8\pi} \int d^2z \left(|\partial
 X^i_{cl}|^2 + |\partial \bar X^i_{cl}|^2 \right),
 \ee 
 and  we have defined
\bea
&& \langle (\partial X^i_{cl})^{{\cal 
 N}_L^i-r}  \,\, (\partial X^i_{qu})^r  (\partial \bar X^i_{cl})^{\bar{\cal N}_L^i-s}  \,\,(\partial \bar
 X^i_{qu})^s\,\, (\bar\partial \bar X^i_{cl})^{\bar{\cal N}_R^i-t}  \,\, (\bar\partial \bar X^i_{qu})^t 
 \rangle_{\sigma^i_{\left(k_1,f_1\right)} \cdots
 \sigma^i_{\left(k_\L,f_\L\right)}} \nonumber \\
&&\quad =  (\partial X^i_{cl})^{{\cal 
 N}_L^i-r}  \,\,  (\partial \bar X^i_{cl})^{\bar{\cal N}_L^i-s}  \,\,(\bar\partial \bar X^i_{cl})^{\bar{\cal N}_R^i-t} 
\times \nonumber \\
&& \quad \quad \int {\mathcal D}X^i_{qu} e^{-S^i_{qu}} (\partial X^i_{qu})^r  \,\,(\partial \bar
 X^i_{qu})^s\,\, (\bar\partial \bar X^i_{qu})^t 
 \sigma^i_{\left(k_1,f_1\right)} \cdots
 \sigma^i_{\left(k_\L,f_\L\right)} \,.
\eea 
We will see that it is not necessary to compute this expression explicitly in order to derive the selection rules.
Indeed, we now use the general form of the correlation function~\eqref{Fqu+Fcl} to deduce that many couplings are forced to be vanishing.

\subsection{3-point couplings}

Let us begin by considering general 3-point couplings.  In this case
we do not have to change pictures, since we have simply the coupling $\langle V_{-1/2}V_{-1/2}V_{-1}\rangle $. Thus, there are no right-moving
oscillators, $\bar \calN_R^i = 0$.  At the same time,  if some of the
participating fields have excited string modes, we do have left-moving
oscillators.  In this case, there are extra selection rules.

\subsubsection{A new rule: ``Rule 5''} \label{S:3ptRule5}

Referring to Eqs.~\eqref{dXcl} and~\eqref{bardXcl}, for 3-point couplings
there exist either non-trivial holomorphic classical solutions ($\sum_\alpha \k_\alpha^i < 2$ where $0 < \k_\alpha^i \leq 1$)
or non-trivial anti-holomorphic solutions ($\sum_\alpha \k_\alpha^i > 1$ where $ 0 \leq \k_\alpha^i <1$)
or neither, but not both. Consider the $i$-th plane and suppose e.g.~that only 
the holomorphic classical solutions are non-trivial.  Then all the terms in 
the sum over $s$ in (\ref{Fqu+Fcl}) vanish apart from the one with $s = \bar{\cal N}_L^i$.  
Next take the sum over $r$.  All these terms are zero apart from the one with  $r = s$ \cite{Hamidi:1986vh},
as follows from the basic OPE \cite{polchinski}
\be
X^{M}(z, \bar z) X^{N}(w, \bar w) \sim -\eta^{MN} \ln |z-w|^2, \qquad M, N = 1, \dots, 10 \,. \label{OPEXX}
\ee
Thus we require ${\cal N}_L^i \geq \bar{\cal N}_L^i$ for the coupling to be non-vanishing, which provides a new selection rule.  Correlation functions that survive this Rule 5 reduce to:
\be
{\mathcal F}^i  =  {\scriptsize \begin{pmatrix} {\cal N}_L^i \\  \bar{\cal N}_L^i \end{pmatrix} }  \sum_{X^i_{cl}} e^{-S^i_{cl}}  (\partial X^i_{cl})^{{\cal 
 N}_L^i- {\bar{\cal N}_L^i}} \,\,\int {\mathcal D}X^i_{qu} e^{-S^i_{qu}}  (\partial X^i_{qu})^{\bar{\cal N}_L^i}  \,\,(\partial \bar X^i_{qu})^{\bar{\cal N}_L^i}\,\, 
 \sigma^i_{\left(k_1,f_1\right)}  \sigma^i_{\left(k_2,f_2\right)}  \sigma^i_{\left(k_3,f_3\right)}  \,.\label{3ptFcl1} 
\ee

Following the same steps for the case that only anti-holomorphic instantons are
allowed, it is easy to see that the correlation functions vanish 
unless $\bar{\cal N}_L^i \geq {\cal N}_L^i$.
Finally, if neither holomorphic nor anti-holomorphic instanton solutions exist, then the correlation functions vanish unless ${\cal N}_L^i = \bar{\cal N}_L^i$.

\subsubsection{Twist invariance}

Assume now for concreteness a coupling for which only holomorphic instantons are allowed, and take the case
$\calN_L^i > \bar \calN_L^i$.  The correlation function reduces as above to Eq.~\eqref{3ptFcl1}, which we rewrite in the shorthand:
\be
{\mathcal F}^i  = \sum_{X^i_{cl}} e^{-S^i_{cl}} (\partial X^i_{cl})^{{\cal 
 N}_L^i-{\bar{\cal N}_L^i}}  \cdot {\cal Z}^i_{qu} \,.\label{3ptFcl} 
\ee
Now, recall the explicit form of the holomorphic
instantons for the 3-point couplings: 
\be
\partial X^i_{cl} = a^i h^i(z) \,. \label{3pthol}
\ee
The coefficient
$a^i=\tilde a^i \nu^i$ is proportional to vectors that are given in Eq.~(\ref{3ptcoset}).  This set of vectors enjoys the $\Z_{\N^i}$ twist symmetry of the 2D plane,  and can thus be arranged
into sets of $\N^i$ vectors with equal length as follows: $\nu^i = \left\{r,
r \omega, r \omega^2, \dots, r \omega^{\N^i-1}\right\}$,  $\left\{2r, 2r
\omega, 2r \omega^2, \dots, 2r \omega^{\N^i-1}\right\}, \dots$, with
$\omega = e^{2\pi i/{\N^i}}$. Contributions to the sum over instantons in
(\ref{3ptFcl}) can be similarly 
arranged: 
\bea
{\mathcal F}^i &=& e^{-\frac{r^2|\tilde a^i|^2}{8\pi}\int
  |h^i(z)|^2} \, (h^i)^{{\calN_L^i}-{\bar{\cal N}_L^i}}\, 
 (r\tilde a^i)^{{\calN_L^i}-{\bar{\cal N}_L^i}}\nonumber \\
&& \times \left( 
1^{{\calN_L^i}-{\bar{\cal N}_L^i}} +\omega^{{\calN_L^i}-{\bar{\cal N}_L^i}}+ \dots +\omega^{(\N^i-1)({\calN_L^i}-{\bar{\cal N}_L^i})}
\right) \cdot {\cal Z}^i_{qu} +\cdots \,. \label{Fdecomp1}
\eea
Using the geometric series we see that the correlation function vanishes unless $\calN_L^i -{\bar{\cal N}_L^i}= 0 \mod \N^i$. This is twist invariance.

\subsubsection{Remembering Rule 4}

Continuing, for a given instanton solution (\ref{3pthol}), when all fields are localized at the same fixed point in the $i$-th plane, the coefficient
$a^i=\tilde a^i \nu^i$ is actually proportional to vectors belonging to 
(a sublattice of) the original torus lattice, which can be read from (\ref{3ptcoset}).  
Suppose the corresponding sublattice has
a $\Z_\K$ automorphism group.  Just as before, the lattice vectors can be arranged
into sets of $\K$ vectors with equal length as follows: $\nu^i = \left\{r,
r \omega, r \omega^2, \dots, r \omega^{\K-1}\right\},$  $\left\{2r, 2r
\omega, 2r \omega^2, \dots, 2r \omega^{\K-1}\right\}, \dots$, now with
$\omega = e^{2\pi i/\K}$. Contributions to the sum over instantons in
(\ref{3ptFcl}) are arranged as before: 
\bea
{\mathcal F}^i &=& e^{-\frac{r^2|\tilde a^i|^2}{8\pi}\int
  |h^i(z)|^2} \, (h^i)^{{\calN_L^i}-{\bar{\cal N}_L^i}}\, 
 (r\tilde a^i)^{{\calN_L^i}-{\bar{\cal N}_L^i}}\nonumber \\
&& \times \left( 
1^{{\calN_L^i}-{\bar{\cal N}_L^i}} +\omega^{{\calN_L^i}-{\bar{\cal N}_L^i}}+ \dots +\omega^{(\K-1)({\calN_L^i}-{\bar{\cal N}_L^i})}
\right) \cdot {\cal Z}^i_{qu} +\cdots \,. \label{Fdecomp}
\eea
Now, let us assume a $\Z_3$ twist in the $i$-th plane.  Twist invariance or R-charge conservation already requires that $\calN_L^i -{\bar{\cal N}_L^i}= 0 \mod 3$ for
a non-vanishing coupling.   The $\Z_3$ planes, however, are 
constructed on a lattice with $\Z_6$ symmetry, namely $SU(3)$ or $G_2$.
As above, it is easy to work out from (\ref{Fdecomp})
that correlation
functions are actually then vanishing unless $\calN_L^i - {\bar{\cal N}_L^i}= 0 \mod 6$.  Analogous statements can of course be made when it is the
anti-holomorphic instantons that are allowed.  
Similar arguments could also apply to a $\Z_2$ twist, on a lattice plane with automorphism group $\Z_4$ or $\Z_6$.  However, it happens that in explicit models,  planes with a $\Z_2$ twist are invariant planes for 3-point couplings, i.e.~at least one field in the coupling is untwisted on this plane.  It then follows that no instanton solutions are allowed there, and Rule 5 imposes the stronger condition, ${\cal N}_L^i - {\bar{\cal N}_L^i}= 0$.

In summary, we have an extra selection rule
whenever the symmetry group of the lattice governing  the couplings is
larger than the point group.  This rule was  first introduced in the literature in
\cite{Hamidi:1986vh,Font:1988tp}, and discussed in \cite{Font:1988nc, Tatsuo}. Before it came to be
forgotten it was known as Rule 4. We can see crystallographically 
that it is relevant for the following types of orbifold planes: 
a $\Z_2$ twist on an $SO(4)$ or $SO(5)$ torus lattice; a $\Z_2$ twist on an $SU(3)$ or $G_2$ 
torus lattice; a $\Z_3$ twist on an $SU(3)$ or $G_2$ torus lattice\footnote{Note that the
sublattices appearing in the various instanton solutions turn out to have the same automorphism symmetry as the original torus lattice, at least for the planes with $\Z_2$ and $\Z_3$ twists.}.

\subsection{Higher order couplings}

We are now ready to discuss general L-point couplings, $\L>3$.  For these
higher order couplings we necessarily have L-3 right-moving
oscillators, distributed amongst the three orbifold planes, from the picture-changing operation.  However, all terms in the sum over 
$r$ in Eq.~\eqref{Fqu+Fcl} vanish apart from\footnote{We are indebted to Robert Richter for discussions on the results of \cite{robert}, which suggested that the original version of this condition was incorrect.  In particular, notice that $\langle \partial X \bar \partial \bar X \rangle$ is finite \cite{Dixon:1986qv}.} $r=s+t$ \cite{Hamidi:1986vh}, as can be derived from the OPE (\ref{OPEXX}) on changing to holomorphic spacetime coordinates.  The
 discussion
of the selection rules is then a straightforward generalization of that for the 3-point
couplings.    

\subsubsection{Rule 5}
Let us first state Rule 5.  There are several different cases.

Consider the $i$-th plane.    If neither the holomorphic nor anti-holomorphic solutions can be non-trivial, then we require 
${\cal N}_L^i = \bar{\cal N}_L^i+\bar{\cal N}_R^i$ and the correlation function takes the form:
\be
{\mathcal F}^i  =  \int {\mathcal D}X^i_{qu} e^{-S^i_{qu}} (\partial X^i_{qu})^{{\cal N}_L^i}  \,\,(\partial \bar
 X^i_{qu})^{\bar{\cal N}_L^i}\,\,(\bar\partial \bar
 X^i_{qu})^{\bar{\cal N}_R^i}\,\, 
 \sigma^i_{\left(k_1,f_1\right)}  \dots  \sigma^i_{\left(k_\L,f_\L\right)} \,.
\ee
If  holomorphic instantons are allowed, but anti-holomorphic instantons are not, then 
${\cal N}_L^i \geq \bar{\cal N}_L^i$, and the coupling is given by:
\bea
{\mathcal F}^i  &=&   \sum_{t=0}^{\text{min}(\bar{\cal N}_R^i, \,{\cal N}_L^i - \bar{\cal N}_L^i)} {\scriptsize \begin{pmatrix} {\cal N}_L^i \\  \bar{\cal N}_L^i +t \end{pmatrix} }  {\scriptsize \begin{pmatrix} \bar{\cal N}_R^i \\  t \end{pmatrix} } \sum_{X^i_{cl}} e^{-S^i_{cl}}  (\partial X^i_{cl})^{{\cal 
 N}_L^i-\bar{\cal  N}_L^i-t} (\bar \partial \bar X^i_{cl})^{\bar{\cal N}_R^i-t} \nonumber \\
&& \qquad \int {\mathcal D}X^i_{qu} e^{-S^i_{qu}} (\partial X^i_{qu})^{\bar{\cal N}_L^i+t}  \,\,(\partial \bar
 X^i_{qu})^{\bar{\cal N}_L^i} \,\,(\bar\partial \bar
 X^i_{qu})^{t}\,\, 
 \sigma^i_{\left(k_1,f_1\right)} \dots  \sigma^i_{\left(k_\L,f_\L\right)}  \,,
\eea
where min stands for the smallest number.  
If instead only anti-holomorphic instantons are allowed, then ${\cal N}_L^i \leq \bar{\cal N}_L^i+\bar{\cal N}_R^i $, and 
\bea
{\mathcal F}^i  &=& {\scriptsize \begin{pmatrix} \bar{\cal N}_L^i \\  {\cal N}_L^i - \bar{\cal N}_R^i \end{pmatrix} }\sum_{X^i_{cl}} e^{-S^i_{cl}}  (\partial \bar X^i_{cl})^{\bar{\cal 
 N}_L^i - {\cal 
 N}_L^i+ \bar{\cal  N}_R^i}  \nonumber \\
&& \qquad \int {\mathcal D}X^i_{qu} e^{-S^i_{qu}} (\partial X^i_{qu})^{{\cal N}_L^i}  \,\, (\partial \bar X^i_{qu})^{{\cal N}_L^i-\bar{\cal N}_R^i}  \,\,(\bar \partial \bar
 X^i_{qu})^{{\cal N}_R^i}\,\, 
 \sigma^i_{\left(k_1,f_1\right)} \dots  \sigma^i_{\left(k_\L,f_\L\right)} \, . \label{onlyantihol}
\eea
If both holomorphic and anti-holomorphic instantons are allowed, 
then all the couplings survive Rule 5, and their correlation functions can be written as:
\bea
&&{\mathcal F}^i = \sum_{s=0}^{\bar{\cal N}_L^i} \sum_{t=0}^{\bar{\cal N}_R^i}{\scriptsize \begin{pmatrix} {\cal N}_L^i \\ s+t \end{pmatrix}}    {\scriptsize\begin{pmatrix} {\bar{\cal N}}_L^i \\ s \end{pmatrix}} {\scriptsize\begin{pmatrix} {\bar{\cal N}}_R^i \\ t \end{pmatrix}}
\sum_{X^i_{cl}} e^{-S^i_{cl}} (\partial X^i_{cl})^{{\cal 
 N}_L^i-s-t}  \,\,   (\partial \bar X^i_{cl})^{\bar{\cal N}_L^i-s}  \,\, (\bar\partial \bar X^i_{cl})^{\bar{\cal N}_R^i-t}  \label{Fqu+Fcl2} \nonumber\\
&& \qquad  \times \int {\mathcal D}X^i_{qu} e^{-S^i_{qu}}  (\partial X^i_{qu})^{s+t} \,\,(\partial \bar X^i_{qu})^{s} \,\,(\bar \partial \bar
 X^i_{qu})^{t}
 \sigma^i_{\left(k_1,f_1\right)} \cdots
 \sigma^i_{\left(k_\L,f_\L\right)}  \,,
\eea
where the non-zero contributions in the sums over $s$ and $t$ satisfy $s+t \leq {\cal N}_L^i$.

Although the above rule is applicable to the couplings including
both the twisted and untwisted sectors, 
we need a remark for the couplings that  include only the 
fully untwisted fields.
If all fields are untwisted, there are no instantonic solutions on 
any plane.  
Then, Rule 5 requires 
${\cal N}_L^i= \bar{\cal N}_L^i +\bar{\cal N}_R^i$ for all of $i=1,2,3$.  Moreover, the masslessness condition implies that charged untwisted matter carry no oscillators.
This implies 
\be
0=\sum_i {\cal N}_L^i -\bar{\cal N}_L^i = \sum_i \bar{\cal N}_R^i =\L-3.
\ee
That is, only the 3-point couplings are allowed among charged
untwisted matter fields, whilst higher order couplings are forbidden \cite{uuu}.
In particular, the $U_1U_2U_3$ couplings are allowed by 
the H-momentum conservation, where we recall that $U_1$, $U_2$, and $U_3$ denote 
the untwisted sectors with the H-momenta, 
$q_{sh}^i=(1,0,0)$, $(0,1,0)$, and $(0,0,1)$, respectively.
Note that untwisted fields are 10D bulk modes.
Dimensional reduction from 10D supersymmetric Yang-Mills theory 
leads to the same result, that is, 
only the $U_1U_2U_3$ couplings are allowed among 
untwisted charged matter couplings, but higher order couplings are 
forbidden by 4D ${\mathcal N}=4$ supersymmetry. 
See however Section \ref{holK} for a way to evade this restriction.

\subsubsection{Twist invariance and Rule 4}
Recall the explicit form for the instanton solutions: 
\bea
&&\partial X^i_{cl}(z) = \sum_{l=1}^{\L-\M^i-1} a^i_l h_l^i (z) =
\sum_{p=1}^{\L-2}\sum_{l=1}^{\L-\M^i-1}\nu_p^i (W^{-1})^{i \,
  p}_{\,\,\,l} h_l^i(z) = 
\sum_{p=1}^{\L-2} \nu_p^i h_p^i(z) ,\label{lambda h}\\
&&\bar \partial X^i_{cl}(\bar z) = \sum_{l'=1}^{\M^i-1} b_{l'}^i
\bar h'^i_{l'}(\bar z) = \sum_{p=1}^{\L-2} \sum_{l'=1}^{\M^i-1} \nu_p^i 
(W^{-1})^{i\,p}_{\,\,\,\L-\M^i-1+l'} \bar{h'}_{l'}^i(\bar z) =
\sum_{p=1}^{\L-2}\nu^i_p \bar{h'}_p^i(\bar z) , \qquad \qquad
\label{lambda h'} 
\eea
plus their complex conjugates.  Here,  the vectors
$\nu_p^i$ belong to particular cosets
of the 
$i$-th torus lattice and $h_p^i, \bar{h'}_p^i$ are defined
in the obvious way via the above equations.  The sets of coset vectors $\left\{\nu_p^i\right\}$ enjoy the $\Z_{\N^i}$ twist symmetry of the $i$-th plane.  For example, for the 4-point coupling, we may write $\nu_1^i = \left\{r_1, r_1 \omega, \dots, r_1 \omega^{\N^i-1}\right\},  \left\{2 r_1, 2 r_1 \omega, \dots, 2 r_1 \omega^{\N^i-1}\right\}, \dots \,,$ and $\nu_2^i = \left\{r_2, r_2 \omega, \dots, r_2 \omega^{\N^i-1}\right\}, \left\{2 r_2, 2 r_2 \omega, \dots, 2 r_2 \omega^{\N^i-1}\right\}, \dots\,,$ with $\omega=e^{2\pi i/\N^i}$.  Moreover, the instanton solutions mediating higher order couplings 
between twisted fields at the same fixed point
are linear combinations of several torus 
sublattice vectors, which enjoy the full torus lattice symmetries\footnote{In principle, different sublattices might each have
different automorphism groups, but for the ${\mathbb Z_2}$ twist and  ${\mathbb Z_3}$ twist planes of interest to us, all sublattices have the same symmetry as the original torus lattice.}.  So, for all twisted fields at the same fixed point we may write $\nu_1^i = \left\{r_1, r_1 \omega, \dots, r_1 \omega^{\K-1}\right\}, \dots\,,$ and $\nu_2^i = \left\{r_2, r_2 \omega, \dots, r_2 \omega^{\K-1}\right\}, \dots\,,$ with $\omega=e^{2\pi i/\K}$ and $\K$ the order of the torus's discrete automorphism group in the given plane.

Let us focus again on the 4-point coupling for concreteness, 
as the general L-point coupling follows straightforwardly.  Assuming that the twisted sectors are such that only holomorphic instantons are allowed, the correlation function can be written as:
\bea
{\mathcal F}^i  &=&  \sum_{X^i_{cl}} (\partial X^i_{cl})^{{\cal 
 N}_L^i-\bar{\cal  N}_L^i-\bar{\cal N}_R^i}  \sum_{t=0}^{{\text{min}(\bar{\cal N}_R^i, \,{\cal N}_L^i - \bar{\cal N}_L^i)} } {\scriptsize \begin{pmatrix} {\cal N}_L^i \\  \bar{\cal N}_L^i +t \end{pmatrix} }  {\scriptsize \begin{pmatrix} \bar{\cal N}_R^i \\  t \end{pmatrix} }\, e^{-S^i_{cl}}\, |\partial X^i_{cl}|^{2(\bar{\cal N}_R^i-t)} \nonumber \\
&& \qquad \int {\mathcal D}X^i_{qu} e^{-S^i_{qu}} (\partial X^i_{qu})^{\bar{\cal N}_L^i+t}  \,\,(\partial \bar
 X^i_{qu})^{\bar{\cal N}_L^i} \,\,(\bar\partial \bar
 X^i_{qu})^{t}\,\, 
 \sigma^i_{\left(k_1,f_1\right)} \dots  \sigma^i_{\left(k_\L,f_\L\right)}  \,. 
\eea
Writing the classical solutions as $\partial X_{cl}^i = \left\{r_1 h_1^i + r_2 h_2^i, r_1 \omega h_1^i + r_2 \omega h_2^i, \dots \right\}$, \\
$\left\{r_1 h_1^i + r_2 \omega h_2^i, r_1 \omega h_1^i + r_2 \omega^2 h_2^i, \dots \right\}, \dots$, it is not hard to see that twist invariance generalised to the 4-point couplings then gives rise to the condition:
\be
{\cal 
 N}_L^i-\bar{\cal  N}_L^i-\bar{\cal N}_R^i= 0 \mod \N^i \,. \label{twistinvfin}
\ee
The same result follows immediately from (\ref{onlyantihol}) for the case where only anti-holomorphic instantons are allowed.  For the general case with both holomorphic and anti-holomorphic instantons, the twist invariance condition (\ref{twistinvfin}) can be similarly derived after factoring out $(\partial X^i_{cl})^{{\cal 
 N}_L^i-\bar{\cal  N}_L^i-\bar{\cal N}_R^i}$, and writing $\partial \bar X_{cl}^i = \left\{r_1 h_1^{'i} + r_2 h_2^{'i}, r_1 \bar \omega h_1^{'i} + r_2 \bar\omega h_2^{'i}, \dots \right\}$, \\
$\left\{r_1 h_1^{'i} + r_2 \bar\omega h_2^{'i}, r_1 \bar\omega h_1^{'i} + r_2 \bar \omega^2 h_2^{'i}, \dots \right\}, \dots$.

We must now apply Rule 4 to all the couplings with twisted fields at the same fixed point that have survived so far.  The derivation of Rule 4 is identical to the derivation of twist invariance with $\N^i$ replaced by $\K$.
Let us therefore only state its consequences.  For a plane with $\Z_3$ twist on a $\Z_6$ lattice, when all fields are at the same fixed point the extra $\Z_2$ symmetry implies ${\cal N}_L^i - \bar{\cal N}_L^i - \bar{\cal N}_R^i = 0 \mod 6$.
For a plane with $\Z_2$ twist on a $\Z_4$ lattice, the extra $\Z_2$ symmetry implies that 
${\cal N}_L^i - \bar{\cal N}_L^i - \bar{\cal N}_R^i = 0 \mod 4$ for non-trivial couplings.  Finally, for a $\Z_2$ twist plane on a $\Z_6$ lattice, non-vanishing correlation functions again meet the condition 
${\cal N}_L^i - \bar{\cal N}_L^i - \bar{\cal N}_R^i = 0 \mod 6$.
Notice thus that for orbifolds with a $\Z_2$ plane, and hence a complex structure, the oscillator couplings depend on the complex structure modulus via the sum over lattice vectors, and for special values of the complex structure modulus an enhanced lattice symmetry forces additional couplings to vanish.
 
Couplings containing untwisted fields
in the plane are also sensitive
to Rule 4, unless all fields are untwisted. This follows from the fact that only the twisted
fields comprised in a coupling contribute non-trivially to the instanton solutions in~\eqref{lambda h},~\eqref{lambda h'},
which inherit the lattice symmetry. 
If all fields are untwisted in the plane, there are no instantonic
solutions there and no additional symmetries emerge.  If all fields are 
fully untwisted, higher order matter couplings are forbidden by Rule 5. Finally, note that, in the case that
twisted fields are localized at different fixed points, the symmetry of the coefficients
of the instantonic solutions is only the one of the twist, so Rule 4 has no effect.

\subsection{On Effective Couplings} \label{holK}

Our interest is in how to identify possible non-vanishing terms, $\Phi^\L$,  in the holomorphic matter superpotential, $W$, of the low energy effective field theory.  We do so by studying the tree-level correlation functions between holomorphic matter fields of the form 
$\langle V_F V_F V_B^{\L-2} \rangle$, where $V_F$ and $V_B$ 
stand for the corresponding fermionic and bosonic vertex operators.  

If a coupling is allowed by our selection rules, it may still be vanishing for other reasons.  Conversely if a given superpotential matter coupling $\langle \psi_1 \psi_2  \phi_3 \dots \phi_\L \rangle$ is forbidden by the selection rules, it may be possible to generate it below some energy scale via an allowed coupling $\langle \psi_1 \psi_2  \phi_3 \dots \phi_\L  s \rangle$, for $s$ some singlet fields,  if the singlet fields acquire vevs at that scale.    

Another way to generate effective couplings is via holomorphic terms in the K\"ahler potential.  For example, in the presence of a $\Z_2$ plane with K\"ahler modulus $T^3$, complex structure modulus $Z$, and two complex Wilson line moduli, $M_1$ and $M_2$, the K\"ahler potential takes the form \cite{AGNT}\footnote{For phenomenological applications of this K\"ahler potential in orbifolds see e.g. \cite{BIM}.}:
\be
\label{Knonholo}
K = -\log\left( (T^3+\bar T^3) (Z+\bar Z) - \frac12 (M_1 + \bar M_2)(\bar M_1 + M_2) \right) \,.
\ee
Expanding the logarithm, we see that the term $e^{K/2} \partial_\alpha\partial_\beta W  \psi^\alpha \psi^\beta$ in the supergravity Lagrangian, corresponds then not only to a coupling $\psi_1 \psi_2 \phi_3 \dots \phi_\L$ from the superpotential, but also to a coupling $\psi_1 \psi_2 \phi_3 \dots \phi_\L \phi_{M_1} \phi_{M_2}$
(though not to the scattering amplitude $\langle V_{\psi_1} V_{\psi_2} V_{\phi_3} \dots V_{\phi_\L} V_{\phi_{M_1}} V_{\phi_{M_2}} \rangle$).  Note that the latter coupling could be interpreted as an effective superpotential term, once the moduli fields have been integrated out.

It is interesting to note that effective superpotential couplings descending from the K\"ahler potential after moduli stabilization are expected to satisfy R-charge conservation, since this corresponds to Lorentz symmetries that survive the orbifold compactification.  In contrast, Rules 4 and 5 emerge from the structure of the tree-level string correlation functions between holomorphic matter fields  $\langle V_ F V_F V_B^{\L-2} \rangle$, which we use to identify the superpotential.  Therefore, the effective couplings obtained after integrating out the moduli may violate Rules 4 and 5, as in the example above.  In order to identify which of these effective couplings are non-vanishing, one must combine the allowed superpotential couplings and the allowed contributions to the Kahler potential.

\section{A concrete example: the $T^6/\Z_{6-II}$ orbifold} \label{S:Z6II}

In this section, we study the L-point couplings in an explicit model,
and in particular the role of Selection 
Rules 4 and 5.  We choose the
$T^6/\Z_{6-II}$ orbifold with twist vector $v^i = \frac16(1,2,-3)$ and underlying torus lattice $G_2 \times SU(3) \times
SO(4)$.
This orbifold has received much attention in the recent
literature, since for certain gauge embeddings and Wilson lines it can
give rise to a massless spectrum containing the MSSM and no
chiral exotics. 
The first, second and third planes have respectively $\Z_6$, $\Z_3$ and $\Z_2$ twist symmetries, but $\Z_6$, $\Z_6$ and $\Z_4$ lattice symmetries.

In principle, since $T^6/\Z_{6-II}$ is a
non-prime orbifold, we must take care to construct the physical
twist-invariant states by taking linear combinations of the basic
twisted states.  However, the calculation for the corresponding
correlation functions reduces in any case to computing auxiliary
correlation functions of the type discussed above.  Moreover, the auxiliary correlations all involve the same sets of quantum numbers, apart from the localization of twisted states in the first plane.  Therefore, if Rules 4 and 5 eliminate one term in the linear combination of auxiliary correlation functions, they eliminate every term.

Below, we will use the H-momentum of the $\theta, \theta^2, \theta^3, \theta^4$ twisted-sector states, which are, respectively, $q_{sh \, \alpha}^i = (\frac16, \frac13, \frac12),(\frac13, \frac23, 0), (\frac12, 0, \frac12), (\frac23, \frac13, 0)$ (the $\theta^5$ sector does not contain positive chiral states). 

\subsection{3-point couplings}

We begin with the 3-point couplings, $\theta^{k_1} \theta^{k_2}
\theta^{k_3}$.  After applying the standard selection rules, the
surviving couplings are of kind
$\theta \theta \theta^4$ and $\theta 
\theta^2 \theta^3$.    
Using the explicit H-momenta for the twisted states given above, it is easy to see that the H-momentum is automatically conserved for these couplings.
Similarly, we can show that the H-momentum is automatically conserved 
for all of the 3-point couplings allowed by the standard selection rule 
in (factorizable)
$T^6/\Z_3$, $T^6/\Z_4$, and $T^6/\Z_{6-I}$  orbifolds.

Consider now the properties of the classical solutions.  Following the discussion around (\ref{dXcl}--\ref{h'}), for the $\theta \theta \theta^4$ coupling, the allowed instanton solutions are the holomorphic ones in the first and second planes.  The fact that in the third plane there is no non-trivial classical solution meets with our intuition,  since the third plane is untwisted for the
$\theta \theta \theta^4$ coupling, implying that there is no
worldsheet instanton contribution to the coupling.  Similarly, for the
$\theta \theta^2 \theta^3$ coupling only the holomorphic instantons in the
first plane are allowed.  Since several of the classical solutions are forced to be vanishing, Rule 5 eliminates various couplings. 

Moreover, for the $\theta \theta \theta^4$  coupling, when we compute explicitly the coefficients $a^1, a^2$ from the global monodromy conditions, we find that the sets of classical solutions have an extra $\Z_2$ symmetry in the second plane, descending from that of the $SU(3)$ torus lattice.  Therefore Rule 4 can forbid some of the couplings that would be mediated by allowed worldsheet instantons in the second plane.

We can now write down the selection rules.  In the cases where there
are no oscillator states involved, only the standard
selection rules are relevant.  In the presence of oscillators, we have in addition 
\begin{itemize} 
\item for $\theta \theta \theta^4$: couplings are non-vanishing
  only if the following three conditions are satisfied (i) ${\mathcal N}_L^1 \geq \bar{\mathcal N}_L^1$ (ii) ${\mathcal N_L}^2 \geq \bar{\mathcal N}_L^2$ (iii) ${\mathcal N}_L^3= \bar{\mathcal N}_L^3$.
Moreover, when all fields are
  at the same fixed point in the second plane,
Rule 4 imposes  ${\mathcal N}_L^2 - \bar{\mathcal N}_L^2 = 0\,\,{\rm
  mod}\,\,6$.  

\item for $\theta \theta^2 \theta^3$: couplings are non-vanishing only
  if (i) ${\mathcal N}_L^1 \geq \bar{\mathcal N}_L^1$ (ii) ${\mathcal N}_L^2=\bar{\mathcal N}_L^2$  (iii)
  ${\mathcal N}_L^3= \bar{\mathcal N}_L^3$.   

\end{itemize}

\subsection{4-point couplings}

Before considering the general case, we discuss in detail the
4-point couplings, $\theta^{\bf k_1} \theta^{\bf k_2} \theta^{\bf k_3}
 \theta^{\bf k_4}$.  There are eleven types of couplings between four
twisted fields, but upon application of the standard selection rules
it turns out that those of interest are
$\theta^4 \theta^4 \theta^3 \theta$, $\theta^3
\theta \theta \theta$ and $\theta^2 \theta^2 \theta \theta$.  
Again, it is easy to check that H-momentum is conserved for these couplings.

As before, we begin by studying the properties of the classical solutions.
For $\theta^4 \theta^4 \theta^3 \theta$, the allowed
instantons turn out to be (i) holomorphic and anti-holomorphic
instantons in the first plane (ii) holomorphic instantons in the
second plane.   For  $\theta^3 \theta \theta \theta$, the allowed instantons are: (i) holomorphic instantons in the first plane
(ii) holomorphic instantons in the second plane (iii)
holomorphic and
anti-holomorphic instantons in the third plane.  Finally, for $\theta^2
\theta^2 \theta \theta$, we have (i) holomorphic instantons in the first plane
(ii) holomorphic and anti-holomorphic instantons in the second
plane. 

Notice that, despite the fact that
the second plane has an untwisted field in the $\theta^4 \theta^4
\theta^3 \theta$ and $\theta^3
\theta \theta \theta$ couplings, it turns out that instanton solutions can still
play a role in the 4-point functions.  Intuitively, we can think of
this as being 
due to the three twisted strings and an untwisted one stretching and interacting to form an intermediate
 worldsheet
instanton state pinned to the three associated fixed points.  The same could not be said for the 3-point couplings, with
two twisted fields and one untwisted field in a given plane, because
if the worldsheet is pinned only to two fixed points it will tend to
collapse.

With the above information in hand, we can write down the additional
selection rules, which apply when there are oscillator states
involved.  The following conditions must be met if the couplings are to be non-vanishing:

\begin{itemize}
\item for $\theta^4 \theta^4 \theta^3 \theta$: Rule 5 imposes (i) ${\cal N}_L^2 \geq
  \bar {\mathcal
N}_L^2$ 
  (ii)  ${\cal N}_L^3 = \bar{\cal N}_L^3 + \bar{\cal N}_R^3$.   Moreover,  Rule 4 imposes
  (iii) ${\cal N}_L^2 - \bar{\cal N}_L^2 - \bar{\cal N}_R^2=  0 \,\, {\rm mod} \,\, 6$, when all three twisted fields are at the same fixed
  point in the second plane.

\item for $\theta^3 \theta \theta \theta$: Rule 5 imposes (i) ${\cal N}_L^1 \geq \bar{\cal N}_L^1$
  (ii) ${\cal N}_L^2 \geq \bar{\cal N}_L^2$.
Moreover, Rule 4 imposes
 (iii) ${\cal N}_L^2 - \bar{\cal N}_L^2 - \bar{\cal N}_R^2=  0 \,\, {\rm mod} \,\, 6$, when all twisted fields
  are at the same fixed point in the second plane 
(iv) ${\cal N}_L^3 - \bar{\cal N}_L^3 - \bar{\cal N}_R^3 = 0 \mod 4$, when all twisted fields are at the same fixed point in the third plane.

\item for $\theta^2 \theta^2 \theta \theta$: Rule 5 imposes (i) ${\cal N}_L^1 \geq \bar{\cal N}_L^1$ 
  (ii)  ${\cal N}_L^3 = \bar{\cal N}_L^3+ \bar{\cal N}_R^3$.  Moreover, Rule 4 imposes 
(iii) ${\cal N}_L^2 - \bar{\cal N}_L^2 - \bar{\cal N}_R^2= 0 \mod 6$, when all 
fields are at the same fixed point in the second plane.

\end{itemize} 

\subsection{General L-point couplings}\label{generalLpt}

Let us write a general L-point coupling  as $(\theta)^{l_1} (\theta^2)^{l_2} (\theta^3)^{l_3} (\theta^4)^{l_4}$ with integers $l_{1,2,3,4}\geq 0$ and $l_1+l_2+l_3+l_4 = \L$.  

We first show that the H-momentum is automatically conserved, once the point group selection rule and R-charge conservation have been imposed.  The point group selection rule constrains the twisted sectors to obey $l_1 + 2 \, l_2 + 3 \, l_3 + 4 \, l_4 = 6\,m$, with $m$ a non-zero natural number.  Meanwhile, using the H-momenta for the various twisted sectors given at the beginning of this section, we can write the H-momentum conservation condition in each plane  as:
\bea
\frac16(l_1 + 2\, l_2 + 3\, l_3 + 4\, l_4) &=& 1 + \bar{\cal N}_R^1 \,,\cr
\frac13(l_1 + 2\, l_2 +  l_4) &=& 1 + \bar{\cal N}_R^2 \,,\cr
\frac12(l_1 +  l_3) &=& 1 + \bar{\cal N}_R^3 \,,
\eea
with integers $\bar{\cal N}_R^i \geq 0$.  We have to check that these conditions can all be satisfied for appropriate choices of $\bar{\cal N}_R^i$.  On applying the point group selection rule, the first condition can immediately be satisfied.  The second condition requires that $l_1+ 2\, l_2 + l_4 = 3\,m$ with $m$ a non-zero natural number, and again the point group selection rules ensures this, unless $l_{1,2,4}=0$.   Similarly, the third condition can be rewritten as $l_1+ l_3 =2\,m$, which is ensured by the point group selection rule, unless $l_{1,3}=0$.     Note, as we saw already in the previous section, the sum of the H-momentum conditions over all planes gives 
$\sum_i \bar{\cal N}_R^i = \L-3$.

Consider now the case where the H-momentum conservation cannot be satisfied because $l_{1,2,4} = 0$.  Since all the fields are then in the third twisted sector, they are all untwisted in the second plane.  The masslessness condition then ensures that these modes have no oscillators in the second plane, so that $\R_\alpha^2 = q_{sh \, \alpha}^2=0$ and $\sum_{\alpha} \R_\alpha^2 = 0$.  These couplings are then excluded by the R-charge conservation law.  Similarly, when $l_{1,3}=0$, all fields are untwisted in the third plane, and it follows that the coupling is excluded by R-conservation in that plane.

Similarly, we can study the H-momentum conservation 
for higher order couplings allowed by the standard selection rules
in (factorizable)
$T^6/\Z_3$, $T^6/\Z_4$, and $T^6/\Z_{6-I}$  orbifolds.

\bigskip

Finally,  we are ready to write down an algorithm to compute when general L-point couplings are allowed.  We recall here that we always consider the positive-chiral states.  

\begin{enumerate}

\item Apply the standard selection rules; gauge invariance, R-charge conservation and the space group selection rule.

\item Use H-momentum conservation to compute how the right-moving oscillators are distributed amongst the three complex planes, 
$\bar{\cal N}_R^i = \sum_{\alpha} q_{sh \,\alpha}^i - 1$.

\item Apply Rule 5.  Check whether holomorphic and anti-holomorphic instantons in the $i$-th plane are allowed.  For non-trivial holomorphic solutions to exist we require 
  $1 + \sum_\alpha (-1 + \k^i_\alpha) < 0$ (where $0 < \k^i_\alpha \leq 1$).  For
  non-trivial anti-holomorphic solutions to exist we require $1 
+
  \sum_\alpha (- \k_\alpha^i) <0$ (where $0 \leq \k^i_\alpha < 1$ ).  If there are neither holomorphic nor anti-holomorphic solutions, then we require 
${\cal N}_L^i = \bar{\cal N}_L^i + \bar{\cal N}_R^i$.  If  holomorphic instantons are allowed, but anti-holomorphic instantons are forbidden, then 
${\cal N}_L^i \geq \bar{\cal N}_L^i$.  If instead only anti-holomorphic instantons are allowed, then ${\cal N}_L^i \leq \bar{\cal N}_L^i+\bar{\cal N}_R^i$.

\item Apply Rule 4 in the second plane: if all twisted fields are at the same fixed point in the second plane, then Rule 4 imposes 
${\cal N}^2_L - \bar{\cal N}_L^2-\bar{\cal N}^2_R  = 0\,\,{\rm mod}\,\,6$.

\item Apply Rule 4 in the third plane: if  all twisted fields are at the same fixed point in the third plane, then Rule 4 imposes 
${\cal N}^3_L  - \bar{\cal N}_L^3 - \bar{\cal N}^3_R = 0\,\,{\rm mod}\,\,4$.

\end{enumerate}

\subsection{Explicit $\Z_{6-II}$ MSSM candidates}
In order to study the phenomenological consequences of the new selection rules,
we proceed now to apply the previous results to identify the allowed couplings 
in the two MSSM candidates studied in~\cite{PRZ} (the defining parameters and matter spectra
are provided in that reference). We identify the admissible couplings between all matter
states of the models up to order 7 in the superpotential, first ignoring the new selection rules and then taking them
into account. 
In this way, we also confirm that each of the rules has a non-trivial effect and eliminate couplings. 
The number of allowed couplings before and after applying the new selection rules  is provided in Table~\ref{tab:couplings}.

We find that around 7\% of the couplings that satisfy the standard selection rules vanish after imposing the new rules.
It is then natural to address the phenomenological impact that this reduction of non-vanishing interactions 
might have. With this aim, we have computed the mass matrices of exotic matter in supersymmetric vacuum configurations 
considering only the standard rules and including the new rules. The mass matrices are (almost) identical 
in both cases and all exotics decouple. We expect that this behaviour is reproduced in all models of the
Mini-Landscape~\cite{Saul}.

\begin{center}
\begin{table}[t]
\centering
\begin{tabular}{|c|c|c||c|c|}
\hline
       & \multicolumn{2}{c||}{Model 1}       & \multicolumn{2}{c|}{Model 2} \\
 order & no rules 4 \& 5 & with rules 4 \& 5 & no rules 4 \& 5 & with rules 4 \& 5 \\
\hline
3 &   116 &    112        &    160 &    152       \\
4 &   144 &    142        &    300 &    290       \\ 
5 &  1446 &   1260(+112)  &   4710 &   4492(+152) \\
6 & 12544 &   8582(+142)  &  55638 &  50194(+290) \\
7 &108796 &  96570(+1372) & 862893 & 833745(+4644)\\
\hline
\end{tabular}
\caption[]{Number of admissible couplings at different orders for the $\Z_{6-II}$  MSSM candidates studied in~\cite{PRZ}.
In the second and fourth columns, we have considered only the standard selection rules discussed in Section~\ref{sec:standardrules}. 
In the third and fifth columns we count the number of couplings that satisfy additionally the new selection rules
4 and 5. In the parentheses, we add the effective couplings that are generated by holomorphic matter terms in the K\"ahler potential once the moduli are integrated out, as discussed in Section \ref{holK}. \label{tab:couplings}} 
\end{table}
\end{center}
\vskip -1.3cm

A positive phenomenological effect of the new rules appears in the proton-decay sector. Proton decay
is produced by the simultaneous presence of the effective operators $Q_i Q_j Q_k L_l$ and 
$\bar{u}_i \bar{u}_j \bar{d}_k \bar{e}_l$ (where the supermultiplets $Q_i$ denote quark doublets, 
$\bar{u}_i,\bar{d}_i$ are up and down-type quark singlets, and $L_i$ represent lepton doublets). These dangerous operators
appear frequently in the effective theories of $\Z_{6-II}$ orbifolds. We find that, in average, only about 30\% of these
couplings (up to order 10 in the superpotential) survive once the new selection rules are applied.

Let us make a final remark. The Mini-Landscape models\footnote{We have verified
that all Mini-Landscape models with two Wilson lines have one pair of vector-like untwisted matter fields.  All but seven Mini-Landscape models with three Wilson lines have one or more vector-like pairs.} used in Table~\ref{tab:couplings} 
have one pair of vector-like untwisted matter fields associated to the $\Z_2$ plane.  As discussed in Section~\ref{holK}, these would provide effective holomorphic matter couplings once the K\"ahler and complex structure
moduli have been integrated out, which we have counted in the parentheses of Table~\ref{tab:couplings}.  
Such couplings could help to address some phenomenological issues~\cite{Kappl:2008ie}.

\section{Discussion}
When building realistic models from string theory, one of the most essential aspects to understand are the couplings within the low energy effective field theory.   The first question to ask is which couplings are non-vanishing, and the answer can be found by applying string coupling selection rules, which are derived from the structure of the corresponding correlation functions.  The correlation functions of interest are tree-level L-point $\langle V_F V_F V_B ... V_B \rangle$, since these correspond to superpotential couplings in the low energy effective field theory.  The rules thus derived can often be understood in terms of symmetries and charge conservation in the effective field theory.

In heterotic orbifold compactifications, the rules usually considered are gauge invariance, R-charge conservation and the space group selection rule.  In this paper we show that there are two additional selection rules, which further restrict the allowed couplings in the superpotential.  
These rules are both relevant when the candidate couplings between twisted fields involve oscillators, which are couplings between excited states and higher order couplings.  

Rule 4 was introduced in the literature in \cite{Hamidi:1986vh,
  Font:1988tp} but has not been applied in recent works.  It is important when the symmetries of the torus lattice, $\Z_{\K^i}$ in a plane $i$, 
are larger than the point group $\Z_{N^i}$.  When all the twisted fields are at
the same fixed point, this additional symmetry is observed in the sum
over instanton solutions that mediate the coupling.   Then, further to
the twist invariance
 ${\cal N}_L^i -\bar{\cal N}_L^i - \bar{\cal N}_R^i =
0 \mod \N^i$, 
this leads to 
${\cal N}_L^i - \bar{\cal N}_L^i - \bar{\cal N}_R^i = 0
\mod \K^i$ for all twisted fields at the same fixed point.   
Notice,
we might also choose to write Rule 4 in terms of the
picture-independent R-charges: $\sum_\alpha \R_\alpha^i = 1 \mod
\K^i$ for all twisted fields at the same fixed point.  Take care,
however, that whereas Rule 4 corresponds to the Lorentz symmetries
that would survive the toroidal compactification, 
the actual R-symmetry in the low energy effective field theory of course corresponds to the Lorentz symmetries that survive the full orbifold compactification.  Indeed, since Rule 4 also depends on the relative distance between the twisted fields in the compact dimensions, it cannot be simply understood as an R-charge conservation in the low energy effective field theory \cite{Font:1988nc}.  Another interesting observation is that the rule eliminates couplings between twisted fields at the same fixed point, which is precisely when we might expect them to be able to interact at order one, field theoretically.  This is somehow similar to the phenomenon that an allowed coupling between twisted fields at the same fixed point may be exponentially suppressed if it involves oscillators.

Rule 5 is another stringy rule, which has not appeared before in the literature.  It arises when the local monodromy conditions and the required convergence properties for the holomorphic and/or antiholomorphic classical solutions, in a given plane $i$, are satisfied only by the trivial solutions, so that worldsheet instanton solutions are not available there to mediate the couplings.  Quantum effects or/and allowed instantons may be sufficient to ensure non-vanishing couplings, but not if the correlation function is proportional to the classical solutions $\partial X_{cl}^i$ and/or $\partial \bar X_{cl}^i$, or their complex conjugates.  This leads to conditions such as 
${\cal N}_L^i \geq \bar{\cal N}_L^i$ if only holomorphic instantons are allowed in the plane $i$. 

We close with a few more observations.  When considering the rules for higher order couplings, a natural question is what are the consequences of picture-changing.  For an $\L$-point coupling of kind $\psi\psi\phi^{\L-2}$, ghost-charge cancellation requires that $\L-3$ picture-changing operators are introduced into the correlation function.  Invariance under picture-changing manifests itself in the fact that the physics -- including the rules -- is invariant under how we choose to distribute the picture-changing amongst the fields
\cite{Dixon:1986qv}.  Note that this seems to make it difficult to express Rule 5 in terms of possible charges carried by each of the participating fields in a coupling; instead the rule depends explicitly on the total number of right-moving oscillators introduced into the correlation function by picture-changing.

Indeed, it remains an important open question whether Rules 4 and 5
correspond to conventional global and/or local symmetries in the low energy
effective field theory, or instead represent intriguing stringy miracles.  We should point out that Rule 5 has not been understood in terms of a symmetry, even at the stringy level.  This is similar to the space group selection rule, which is due to the boundary conditions of the interacting strings, although the latter can at least be partially  understood in terms of various $\Z_N$-type global symmetries \cite{buchmullerII,tatsuo-stuart,jonas}. 

In many models, R-symmetries and other discrete symmetries are 
anomalous \cite{Araki:2008ek}.
It would be important to study how those anomalies affect 
 Rule 4 as well as Rule 5.
We will study these aspects elsewhere.

Another interesting problem is the extension of the rules to the case
of non-factorizable orbifolds.  
Those include $T^6/\Z_7$, $T^6/\Z_{8-I}$, $T^6/\Z_{8-II}$,  
$T^6/\Z_{12-I}$ and $T^6/\Z_{12-II}$ orbifolds 
\cite{Katsuki:1989bf,Kobayashi:1991rp}.  
In addition, even $T^6/\Z_3$, $T^6/\Z_4$, $T^6/\Z_{6-I}$ and
$T^6/\Z_{6-II}$ orbifolds as well as $T^6/(\Z_M \times \Z_N)$
orbifolds are also realized as non-factorizable orbifolds 
\cite{Katsuki:1989bf,Kobayashi:1991rp,Faraggi:2006bs,Forste:2006wq,Takahashi:2007qc,Ploger:2007iq}.
In fact, even the R-charge conservation law has not yet been understood for non-factorizable orbifolds \cite{saulsthesis}.

Our focus in the present paper has been in selection rules for superpotential couplings.  The ultimate objective would be a derivation of the full low energy effective field theory describing the orbifold compactification, in terms of the superpotential, K\"ahler potential, gauge kinetic functions and Fayet-Iliopolous terms.  Much less is known about the K\"ahler potential, but it can be derived by computing four boson scattering amplitudes between matter fields and moduli fields \cite{DKL}, and can include some contributions where all the matter fields are holomorphic.  After moduli stabilization, such terms would lead to new couplings in an effective superpotential, relevant for phenomenology.  Thus it would be essential to study further the K\"ahler potential.

Finally, we have briefly illustrated the implementation of the new rules in some $\Z_{6-II}$ orbifold MSSM candidates, 
including the known holomorphic matter contributions to the K\"ahler potential.  
  Couplings are crucial to understanding the dynamics of such models, for instance the decoupling of exotics, as well as for example the quark and lepton masses.  Our initial results indicate that the heterotic orbifold Mini-Landscape \cite{Saul} continues to provide promising phenomenological models.  The new selection rules must now be implemented in all such orbifold studies.

\vspace{1cm}

\section*{Acknowledgements}
We would like to thank M.~Bianchi, N.~Cabo-Bizet,  O.~Lebedev,  C.~L\"udeling, D.~Mayorga Pe\~na, H.~P.~ Nilles, N.~Pagani, M.~Ratz, R.~Richter, G.~Tasinato, C.~Vafa and P.~Vaudrevange for helpful discussions.  We are especially grateful to R.~Richter for indicating an error in our original version of Rule 5.  
T.~K. is supported in part by the Grant-in-Aid for 
Scientific Research No.~20540266  and the Grant-in-Aid for the Global COE 
Program ``The Next Generation of Physics, Spun from Universality and 
Emergence" from the Ministry of Education, Culture, Sports, Science and 
Technology of Japan.
S.~L.~P is supported by the
G\"{o}ran Gustafsson Foundation.
I.~Z.~was supported by the DFG cluster of excellence
Origin and Structure of the Universe, the SFB-Tansregio TR33
``The Dark Universe" (Deutsche Forschungsgemeinschaft) and
the European Union 7th network program ``Unification in the
LHC era" (PITN-GA-2009-237920). S.~R.-S. was partially 
supported by CONACyT project 82291 and DGAPA project IA101811.

\end{document}